\documentclass{article}
\pdfoutput=1
\usepackage{dcase2020,amsmath,graphicx,url,times,booktabs, tabularx, amssymb, standalone}

\title{Multi-task Regularization Based on Infrequent Classes for Audio Captioning}
\name{Emre \c{C}ak\i{}r, Konstantinos Drossos, and Tuomas Virtanen\thanks{The authors would like to acknowledge CSC-IT Center for Science, Finland, for computational resources.}}
\address{Audio Research Group\\
Tampere University\\
Tampere, Finland\\
\{firstname.lastname\}@tuni.fi}

\begin{document}

\ninept
\maketitle

\begin{sloppy}

\begin{abstract}
Audio captioning is a multi-modal task, focusing on using natural language for describing the contents of general audio. Most audio captioning methods are based on deep neural networks, employing an encoder-decoder scheme and a dataset with audio clips and corresponding natural language descriptions (i.e. captions). A significant challenge for audio captioning is the distribution of words in the captions: some words are very frequent but acoustically non-informative, i.e. the function words (e.g. ``a'', ``the''), and other words are infrequent but informative, i.e. the content words (e.g. adjectives, nouns). In this paper we propose two methods to mitigate this class imbalance problem. First, in an autoencoder setting for audio captioning, we weigh each word's contribution to the training loss inversely proportional to its number of occurrences in the whole dataset. Secondly, in addition to multi-class, word-level audio captioning task, we define a multi-label side task based on clip-level content word detection by training a separate decoder. We use the loss from the second task to regularize the jointly trained encoder for the audio captioning task. We evaluate our method using Clotho, a recently published, wide-scale audio captioning dataset, and our results show an increase of 37\% relative improvement with SPIDEr metric over the baseline method.
\end{abstract}


\section{Introduction}\label{sec:introduction}
Audio captioning is the novel task of automatically generating textual descriptions (i.e. captions) of the contents of general audio recordings~\cite{drossos:2020:icassp,lipping:2019:dcase}. Audio captioning started in 2017~\cite{drossos:2017:waspaa}, and it can be considered as an inter-modal translation task, where the humanly perceived information in the audio signal is translated to text. For example, given an audio recording, a caption can be ``\emph{the wind blows while cars are passing by}'' or ``\emph{a man alternates between talking and flapping a piece of cloth in the air three times}''\footnote{Actual captions from Clotho dataset.}. 

Existing audio captioning methods are deep neural networks (DNNs) based, mostly employing the sequence-to-sequence paradigm. An encoder gets as an input the audio sequence, processes it, and outputs a sequence of learned feature vectors. Then, the output sequence of the encoder is aligned with the targeted output sequence of the decoder, typically by two alternative methods. The first, is the one proposed in~\cite{cho:2014:learning}, where the encoder outputs a fixed length vector, and this vector is used as an input to the decoder and for every time-step of the output sequence. The second method for sequence alignment, is through the attention mechanism proposed in~\cite{bahdanau:2014:iclr}, where for each time-step of the output sequence, the alignment mechanism calculates a weighted sum of the output sequence of the encoder, conditioned on the state of the decoder. For example, in~\cite{drossos:2017:waspaa} the method uses an encoder-decoder scheme, utilizing a multi-layered and recurrent neural network (RNN) based encoder and an RNN-based decoder. The encoder gets as an input the audio signal, and its output is processed by the attention mechanism presented in~\cite{bahdanau:2014:iclr}. The output of the attention mechanism is used as an input to an RNN-based decoder, followed by a classifier which outputs the predicted words at each time-step of the output sequence. Study~\cite{wu:2019:icassp} presents another method, where the input sequence is encoded to a fixed length vector, through an RNN-based encoder and by a time-averaging of the output of the encoder. Then, the fixed length vector is used as an input to an RNN-based decoder, for every time-step of the output sequence, similarly to~\cite{cho:2014:learning}. Again, a classifier predicts the output words. Finally, the work in~\cite{kim:2019:acl} presents an approach where a VGGish encoder is used to process the input audio sequence. The output of the VGGish-based encoder is processed by an attention mechanism, and a sub-sequence RNN-based decoder followed by a classifier, outputs the predicted words. Finally, the method in~\cite{drossos:2020:icassp} is also employed for establishing initial results for the audio captioning dataset called Clotho. Clotho is a novel audio captioning dataset, employing around 5000 clips with five captions for each audio clip, amounting to a total of around 25 000 audio clips and caption examples. Clotho is build with emphasis on diversity and robustness, utilizing established good practises for dataset creation from the machine translation and image captioning communities~\cite{drossos:2020:icassp,lipping:2019:dcase}. Clotho offers captions that are sanitized from speech transcription, typos, and named entities, provides splits with no hapax-legomena (i.e. words appearing only once in a split)~\cite{popescu:2008:jql} and it is used in the audio captioning task of the DCASE 2020 challenge\footnote{\label{fn:dcase-task}\url{http://dcase.community/challenge2020/task-automatic-audio-captioning}}. 

Though, in many natural language-based datasets (like Clotho) it is observed that there is a typical class imbalance~\cite{Padurariu:2019:pcs,buda2018systematic}. The function words, i.e. the articles (e.g. ``a''/``an'', ``the''), prepositions (e.g. ``in'', ``over'', ``from'', ``about''), and conjunctions (e.g. ``and'', ``or'', ``the'', ``until'') appear in overwhelming amounts compared to the other words, i.e. the content words. The frequency of appearance of content words, most likely will cause the common machine learning optimization methods, such as gradient descent, to overfit to them, since they are the ones that factor the most to the loss function. This is especially undesirable for audio captioning in two major ways. Function words most often do not possess any information about the audio content, which makes it even harder to map the acoustic features to these most common words, i.e. classes. In addition, the class imbalance between function and content words, prevents the learning of the acoustically more informative, content words, since they contribute less to the total learning loss. On the other hand, a valid caption generated by an automatic audio captioning system must anyway include the function words in the appropriate places to be grammatically correct.

To tackle the above, we draw inspiration from traditional natural language processing techniques. Specifically, we consider the cases where the class imbalance between function and content words is treated with employing weights of the loss for the words~\cite{Padurariu:2019:pcs,sheng:2006:aaai}, and we propose a novel regularizing method for the encoder that process the audio sequence, employing a data pre-processing and a multi-task learning set-up. We first identify the function and content words. Then, additionally to the task of predicting the proper sequence of words (i.e. caption) for a given audio input, we utilize an extra learning signal for the encoder. This signal emerges from an extra decoder followed by a classifier, which try to predict the content words for the corresponding input. 

The rest of the paper is organized as follows. In Section~\ref{sec:method} is our proposed method and Section~\ref{sec:evaluation} describes the followed evaluation procedure. The obtained results and their discussion are in Section~\ref{sec:results}. Section~\ref{sec:conclusions} concludes the paper and proposes future research directions. 

\section{Proposed method}\label{sec:method}
The proposed method consists of two stages: feature and target extraction, and deep learning based sequence-to-sequence classifier. Given an audio recording, our method first extracts the acoustic features, and then uses a recurrent neural network based autoencoder with two separate decoders to generate an audio content description as a sentence. While the system is trained with both caption decoder and content word decoder, only caption decoder is used to obtain the generated content description. The system overview is given in Figure~\ref{fig:overview}.

The proposed method is based on the baseline system for audio captioning task of the DCASE 2020 challenge\textsuperscript{2}, and includes several extensions to this work. In order to alleviate the class imbalance problem on the learning of acoustically informative words, we propose two main extensions: multi-task regularization based on content words, and loss weighting based on word frequency. For the rest of the paper, these methods are referred as CWR-CAPS (content word regularized captioning system) and CWR-WL-CAPS (content word regularization with weighted loss captioning system). The open-source code repository for this work, written as an extension to official challenge baseline system, is available online\footnote{\label{fn:github-repo}\url{https://github.com/emrcak/dcase-2020-baseline/tree/sed_caps}}.

\begin{figure*}[!t]
    \centering
    \includegraphics[width=0.85\linewidth]{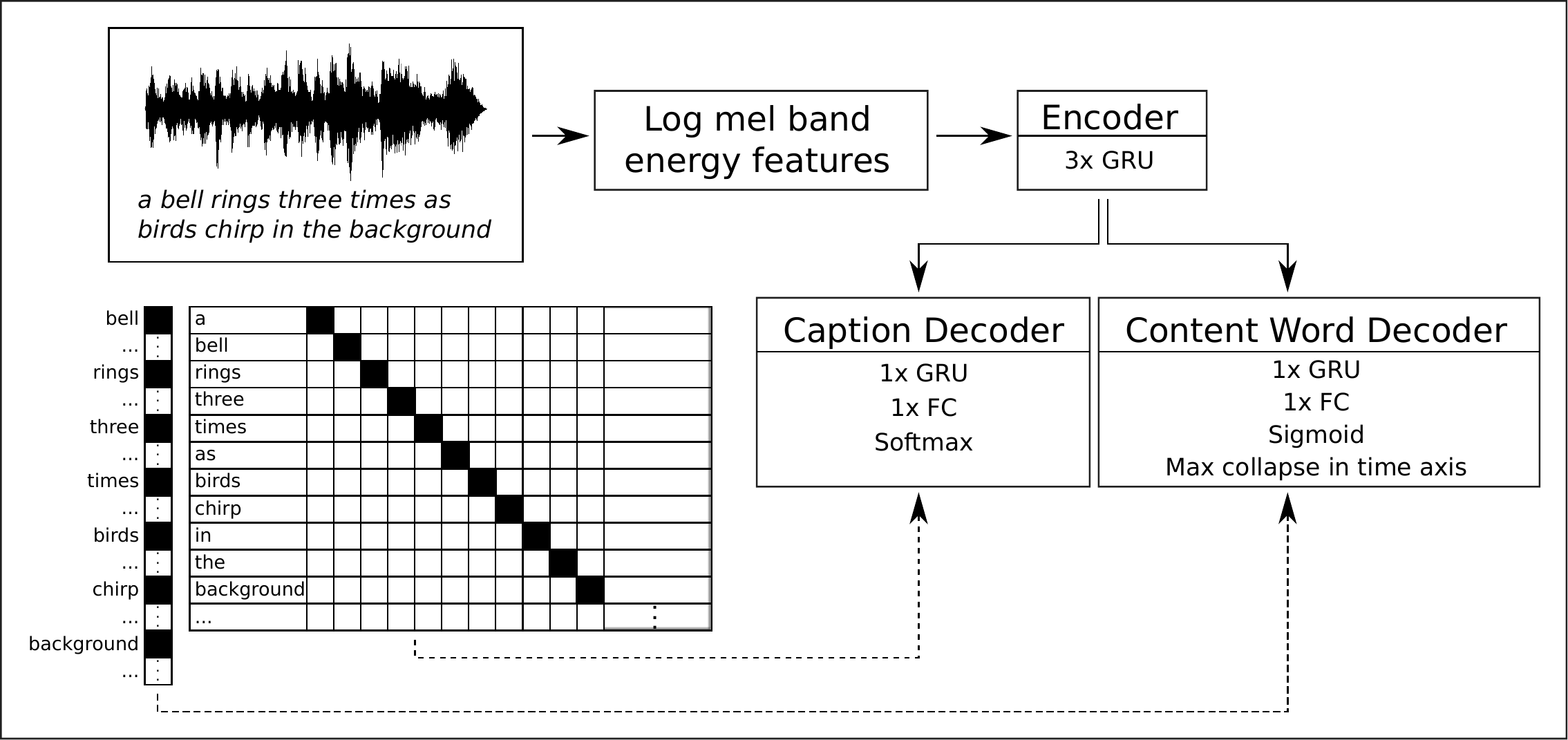}
    \caption{System overview.}
    \label{fig:overview}
\end{figure*}

\subsection{Feature and target extraction}\label{subsec:data}
We use log mel band energies as acoustic features. Since Clotho dataset comes with varying length audio recordings, zero-padding is applied at the end for the shorter recordings in each batch. As a result, the acoustic features $\mathbf{X} \in \mathbb{R}^{N\times T}$ for each recording are obtained, where $N$ is the number of bands, and $T$ is the maximum number of frames in a recording for a given batch.

Each caption is pre-processed by making all the words lower-case, removing the punctuation and adding start- and end-of-sequence tokens ([SOS] and [EOS]). The target outputs for captions of a recording is a matrix $\mathbf{Y} \in \mathbb{R}^{K\times T'}$, where $K$ is the number of unique words in the dataset including [SOS] and [EOS], and $T'$ is the length of the longest caption target output vector in a batch. Each column of $\mathbf{Y}$ is a one-hot vector representing the index of a word in the caption at each timestep. Similar with the input features, the target output length is varying among the recordings, therefore the shorter target outputs in each batch are padded with [EOS] tokens. 

\subsubsection{Content word extraction}
We define a second set of the target outputs for audio captioning, namely \textit{content words}, to be used to introduce regularization over the encoder outputs. The content words are defined as the set of words in the given dataset of captions, excluding the prepositions, articles, conjunctions and auxiliary verbs. The list of content words used in this work can be found in the open-source code repository\textsuperscript{3}. As a result, for each given caption, we obtain a multi-label encoded binary content word vector $\mathbf{y}' \in \mathbb{R}^{K'}$. If the $i^{th}$ content word is present in the caption, then $\mathbf{y}'_{i}$ is set to 1, and 0 vice versa.

\subsection{Sequence-to-sequence classifier}
The proposed method is a sequence-to-sequence deep learning classifier with two sets of target outputs: captions and content words. The input to the system is log mel band energy features. This input is fed to an encoder block which consists of bidirectional Gated Recurrent Unit (GRU)~\cite{cho2014properties} layers. Dropout~\cite{srivastava2014dropout} is applied after each GRU layer.

The output of the encoder is then fed to two separate decoder branches, namely caption decoder and content word decoder. The difference between captions and content words, and how the content words are obtained are explained in Section~\ref{subsec:data}. Both decoder blocks include a single unidirectional GRU layer, and a fully connected (FC) layer. The FC layers in both decoder blocks apply the same set of weights over the RNN outputs at each timestep.

The differences between the caption and content word decoder processes are as follows. The main difference is the nonlinearity applied to the weighted outputs. In the case of caption decoder, the FC layer nonlinearity is softmax, whereas for the content word decoder, it is sigmoid function to allow multiple content word outputs being detected for the same input. Another difference is that the final outputs for the content word decoder are collapsed in time axis by taking the maximum value over time, in order to obtain a single probability vector for the whole recording. The caption decoder outputs are also treated as probabilities at the frame level, where each column represents the probabilities of the words at a given frame (timestep). During inference, the caption decoder output is determined as the word with the highest probability at each time step.

\subsubsection{Training}
\label{subsubsec:optimization}
The sequence-to-sequence classifier is trained using Adam gradient optimizer~\cite{adamKeras}. The upper boundary of the squared norm of the gradients is selected as 1 to prevent exploding gradients. The classifier is trained using a patience scheme, where the training is aborted if the SPIDEr metric for the evaluation dataset does not improve for ceratin number of epochs. As the final model, we use the model from the epoch with the best validation SPIDEr score.

Weighted non-negative log likelihood and cross-entropy loss are used as objective loss functions for caption decoder and content word decoder outputs, respectively. The purpose of using the additional loss from the content words is to regularize the encoder to produce intermediate representations that contain more information on the content words. Our empirical analysis show that the magnitude of loss corresponding to the content words is consistently about 10\% of the captioning loss over the training. Therefore, this additional loss does not dominate the whole training and can indeed be seen as acting as a regularizer over the encoder. 

The class weights for the non-negative log likelihood loss of the caption decoder output are determined as inversely proportional to the amount of each word's occurrences in the development dataset. This leads the classifier to avoid overfitting on more common but less informative words, due to their smaller weight on the total loss function. This method also provides a better matching with some of the commonly used captioning evaluation metrics such as CIDEr~\cite{vedantam2015cider}, which uses Term Frequency Inverse Document Frequency~(TF-IDF)~\cite{robertson2004understanding} weighting that puts more emphasis on the detection of the less common words. 

\subsubsection{Other changes to baseline method}
Apart from the proposed content word based regularization and weighted loss schemes, there are a few additional changes made to baseline system. For the baseline system, the input features from the shorter (in time) recordings are padded with zero vectors for the beginning timesteps to have an equal sized feature matrix between the examples in a batch. In order to better match the target outputs being padded at the end, we move the input feature padding also to the end. In addition, validation SPIDEr score based early stopping is added to the baseline system. Also, the gradients are reset after processing each batch (this was initially missing from the baseline system - hence from baseline results -, but later added to baseline code repository).

\section{Evaluation}\label{sec:evaluation}
\subsection{Dataset}
In correspondence with DCASE 2020 challenge task on audio captioning (task 6), Clotho~\cite{Drossos_2020_icassp} dataset is used for development and evaluation. Clotho consists of 15 to 30 seconds long recordings collected from FreeSound platform~\footnote{\label{fn:freesound}\url{https://freesound.org/}}, and each recording is annotated with five different captions using crowd-sourcing. In this work, development split of Clotho is used for training the systems, and the performance is evaluated using the evaluation split.

\subsection{Evaluation Metrics}
For the assessment of the performance of our method, we employ the proposed metrics from the audio captioning task at DCASE 2020 challenge\textsuperscript{2}. These metrics can be divided in two categories. Firstly there are the machine translation metrics, which are $\text{BLEU}_{n}$~\cite{papineni:2002:bleu}, $\text{ROUGE}_{\text{L}}$~\cite{lin:2004:rouge}, and METEOR~\cite{lavie:2007:meteor}. $\text{BLEU}_{n}$ calculates a weighted geometric mean of the precision of $n$-grams (typically $n\in\{1, 2, 3, 4\}$) between predicted and ground truth captions, $\text{ROUGE}_{\text{L}}$ calculates an F-measure using the longest common sub-sequence (also between predicted and ground truth captions), and METEOR is based on a harmonic mean of the precision and recall of segments, from the predicted and ground truth captions. Then, there are the captioning metrics which are the CIDEr~\cite{vedantam:2015:cider}, SPICE~\cite{anderson:2016:spice}, and SPIDEr~\cite{2017:liu:iccv}. CIDEr uses a weighted sum of the cosine similarity of $n$-grams, between the predicted and ground truth captions, and SPICE measures how well the predicted caption recovered objects, scenes, and relationships of those, according to the ground truth caption. SPIDEr is the mean of CIDEr and SPICE, exploiting the both of each metrics~\cite{2017:liu:iccv}. 

\subsection{Hyperparameters}
The specific hyperparameters used in this work are as follows. The feature extraction and the model architecture hyperparameters are kept the same with the baseline method for better comparability. The number of log mel bands for feature extraction is selected as 64, and Hamming window of 46 ms length with 50\% overlap is used for frame division. the total number of content words is 88. For the encoder, we use three bidirectional GRU layers with 512 units each. The dropout probability used in the encoder is 0.25. For the decoder, we use one GRU layer with 512 units. For the autoencoder classifier training, the batch size is selected as 32 and the Adam learning rate is selected as $10^{-4}$. The maximum number of training epochs is set to 300, with 100 patience epochs before aborting.  

\section{Results and discussion}\label{sec:results}
The performance results for CWR-CAPS and CWR-WL-CAPS with the DCASE 2020 challenge official metrics are given in Table~\ref{tab:results}. CWR-WL-CAPS method offers 37\% relative increase on SPIDEr compared to baseline, and also performs better on other metrics. Moreover, comparing CWR-CAPS and CWR-WL-CAPS, there is a considerable benefit for using weighted loss for the caption outputs. The benefit is more evident in CIDEr metric compared to SPICE. This is also consistent with the theoretical expectations, due to TF-IDF weighting in CIDEr calculation (as mentioned in Section~\ref{subsubsec:optimization}).
\begin{table}
    \centering
    \begin{tabular}{l| c c c}
        \toprule
        Metric & Baseline & CWR-CAPS & CWR-WL-CAPS  \\
        \midrule
        $\text{B}_{1}$ & 38.9 & 39.0 & 40.9 \\
        $\text{B}_{2}$ & 13.6 & 14.3 & 15.6  \\
        $\text{B}_{3}$ & 5.5 & 6.3 & 7.3 \\
        $\text{B}_{4}$ & 1.5 & 2.4 & 3.0 \\
        $\text{R}$ & 26.2 & 27.0 & 27.8  \\
        M & 8.4 & 8.5 & 8.8 \\
        \midrule
        CIDEr & 7.4 & 8.9 & \textbf{10.7} \\
        SPICE & 3.3 & 3.6 & \textbf{4.0} \\
        SPIDEr & 5.4 & 6.3 & \textbf{7.4} \\
        \bottomrule
    \end{tabular}
    \caption{Percentage results for baseline, CWR-CAPS and CWR-WL-CAPS. $B_N$ stands for BLEU, $R$ is for ROUGE, and $M$ is for METEOR.}
    \label{tab:results}
\end{table}

While both methods perform better than the baseline, the produced captions still mostly lack the structure of a grammatically valid sentence. Even though the non content word contribution to the objective loss is decreased significantly, words such as \textit{is}, \textit{are}, and \textit{and} appear repetitively towards the end of many of the produced captions. This can be attributed to the fact that both caption and content word autoencoders aim to map acoustic features to the words, without e.g. a language-model based prior. Longer-term temporal modeling of the captions can be improved with a language model, trained with the given captions and also with external text material, which would be then used together with the autoencoder acoustic model, and the outputs would be produced using e.g. beam search algorithm~\cite{tillmann2003word}. We currently consider this approach as the future work for this task.

\section{Conclusions}\label{sec:conclusions}
In this paper, we propose two methods for automated audio captioning. These methods are based on content word regularization and weighted objective loss, both using recurrent neural network based autoencoder. This work is evaluated in the framework of DCASE 2020 challenge task on audio captioning, and both proposed methods provide a considerable boost over the baseline results of the challenge. Still, the produced captions mostly lack the correct English grammatical structure, and addressing this problem using external language models is planned as future work.

\bibliographystyle{IEEEtran}
\bibliography{refs}
\end{sloppy}
\end{document}